\documentclass{emulateapj}
\usepackage{ulem}
\slugcomment{draft v2}

\shorttitle{Long-lasting BH Jets in Short GRBs}
\shortauthors{Kisaka \& Ioka}

\begin{document}


\title{Long-Lasting Black-Hole Jets in Short Gamma-Ray Bursts}


\author{Shota Kisaka\altaffilmark{1}}
\email{kisaka@post.kek.jp}
\author{Kunihito Ioka\altaffilmark{1,2}}
\email{kunihito.ioka@kek.jp}


\altaffiltext{1}{Theory Center, Institute of Particle and Nuclear Studies, KEK, Tsukuba 305-0801, Japan}
\altaffiltext{2}{Department of Particle and Nuclear Physics, SOKENDAI (the Graduate University for Advanced Studies), Tsukuba 305-0801, Japan}


\begin{abstract}

Whether a short gamma-ray burst (GRB) is caused by a black hole (BH) or a neutron star (NS) after the merger of a NS binary is a crucial problem. 
We propose a BH model that explains short GRBs with long-lasting activities such as extended emission and plateau emission up to $\sim10000$ s. 
To extract the BH rotational energy, the topological evolution of the magnetic field should accompany the mass ejection, mass fallback, and magnetic field reconnection.
The observations suggest the magnetic field decay from $\sim10^{14}$ G to $\sim10^{13} - 10^{11}$ G at the BH, bounded below by the pre-merger strength and kept constant while the luminosity is constant, and the fallback mass of $\sim10^{-4} - 10^{-2} M_{\odot}$, comparable to the ejecta mass implied by the macronova (or kilonova) in GRB 130603B. 
The BH model has implications for gravitational waves and the equation of state of NS matter.

\end{abstract}





\section{INTRODUCTION}

A black hole (BH) or a neutron star (NS)? 
This is the fundamental question about the central engine of short gamma-ray bursts \citep[GRBs; e.g., ][]{ZM04, B14}.
A leading model for the short GRBs is a binary NS merger.
A merger produces a massive NS, which may or may not collapse to a BH depending on the equation of state of NS matter \citep{BBM13}.
In a BH-NS binary merger, the central engine would be the BH.
Obviously, there is a significant difference in the jet formation between a BH and NS.
The central engine also affects the gravitational wave (GW) signals \citep{Hot+11} and the electromagnetic counterparts \citep[e.g., ][]{NP11, Gao+13, TKI14, KIT15}.
Detection of these signals will be soon realized by laser interferometers such as Advanced LIGO, Advanced VIRGO and KAGRA. 

X-ray observations suggest long-lasting activities of the central engines of short GRBs.
Following the prompt GRB for $\sim10^{-2}-1$ s, some events show extended emission \citep[$\sim10^2$ s;][]{Bar+05}, which lasts much longer than the typical accretion 
\footnote{We define the extended emission as the emission for $\sim10^2$ s, which also includes the plateau component analyzed by \citet{Row+13} and \citet{Lu+15}.
}.
Furthermore, some of events accompany even a longer activity ($\sim10^3-10^4$ s; hereafter we call this plateau emission) \citep{GOWR13, GOW14} 
\footnote{The plateau components with long timescale ($\sim10^3-10^4$ s) are also identified by \citet{Row+13}. Note that the plateau emission would be sometimes hidden by the afterglow emission or below the detection limit. These events may correspond to the ``no breaks'' in \citet{Row+13} and ``no plateau samples'' in \citet{Lu+15}.}. 
The sharp drop of the light curve is produced only by the activity of the central engine \citep{IKZ05}.
In this {\it Letter}, we consider the samples in \citet{GOW14}. 

To explain the long-lasting activities, a highly magnetized and rapidly spinning NS model (magnetar model) is considered \citep[e.g., ][]{MQT08}. 
This comes from the fact that the spin-down timescale is $\sim10^3$ s for a NS with the dipole magnetic field $\sim10^{15}$ G and the initial rotation period $\sim1$ ms. %\citep[e.g., ][]{U92}.
The spin-down luminosity is also comparable to the observed one within a reasonable range of the parameters 
\footnote{The prompt GRB jet should be launched before the outflow of the long-lasting activities because the outflow becomes too thick for the jet to propagate keeping the prompt timescale $\sim 0.1$ s \citep{Nag+14} unless the outflow is unreasonably cold \citep{RK14,CS14}.
}.

In the case of the BH engine, fallback accretion is considered to produce the long-lasting activities \citep[e.g., ][]{LR07}. 
In fact, the released energy is enough to power the observed luminosity.
However, the accretion rate of the fallback matter follows a single power-law \citep[see Equation (\ref{sec2:dotM}); ][]{R07}. 
Then, the resulting activity does not seem to explain the characteristic timescales in the light curve.

In this {\it Letter}, we present a BH model that reproduces the long-lasting activities of short GRBs.
The point is the topological evolution of the magnetic field associated with the mass ejection and fallback processes. 
We show that the mass ejection, mass fallback, and magnetic field decay via reconnection are all inevitable events for the jet formation by Blandford-Znajek (BZ) process \citep{BZ77}.
We find that the magnetic field changes to $\sim10^{14}$ G, $\sim10^{13}$ G, and $\sim10^{12}$ G in the prompt, extended and plateau emission, respectively. 
The required mass of the fallback matter is $\sim10^{-4}-10^{-2}M_{\odot}$, which is consistent with the recent numerical simulations \citep[e.g., ][]{Hot+13b, KIS13, Kyu+15} as well as the observed infrared excess (the so-called macronova \footnote{We use the term ``macronova'' as a transient with a  binary NS merger, especially thermal radiation from the merger ejecta, including an engine-powered macronova \citep{KIT15}.} or kilonova) associated with short GRB 130603B \citep[e.g., ][]{Tan+13, BFC13, TNI14, KIT15}. 
We introduce our BH model in Section 2.
In Section 3, we compare the theoretical light curves with observations. 
Discussions follow in Section 4.

\section{BLACK HOLE MODEL}

 \begin{figure*}
  \begin{center}
   \includegraphics[width=165mm]{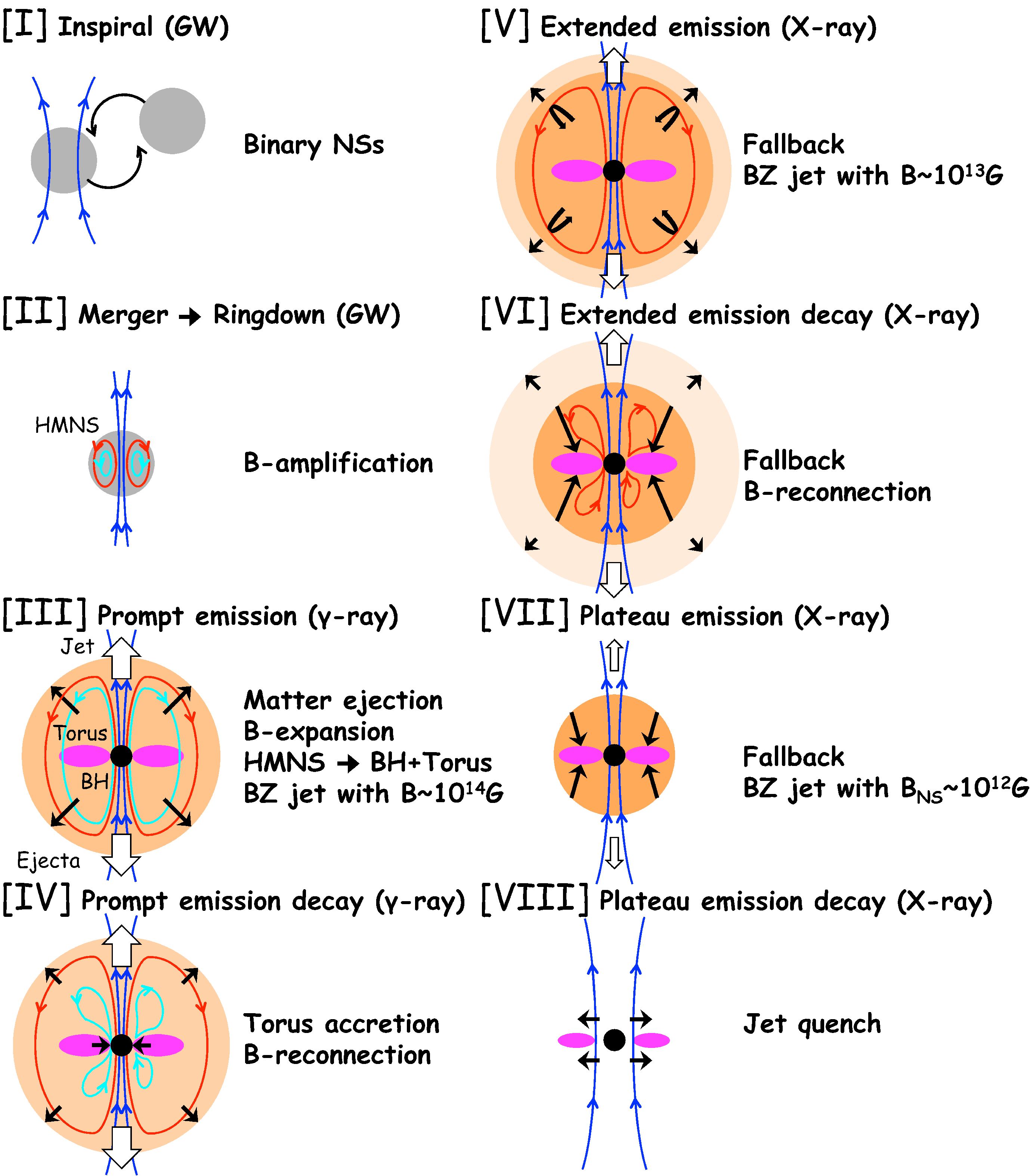}
   \caption{Schematic pictures of our BH model for short GRBs. See Section 2 for details.}
   \label{figure:model}
  \end{center}
 \end{figure*}

In our BH model, a BH is the central engine of short GRBs including long-lasting activities such as extended emission and plateau emission. 
We consider the BZ process for the BH to launch a relativistic jet \citep{BZ77} since the neutrino-neutrino annihilation is not effective at late time ($>1-10$ s).
Our model is based on general topological consideration in Figure~\ref{figure:model}, without resorting to specific processes such as radiation mechanisms.

There are three key ingredients for the BZ process, (i) rotation of the BH, (ii) a magnetic field strength on the BH, and (iii) large-scale, poloidal configuration of the magnetic field, which means that the characteristic size of the poloidal field on the BH is much greater than the outer light cylinder \citep{BHK08}. 
We consider the BH with mass $M_{\rm BH}$, a spin parameter $a=Jc/GM_{\rm BH}$ and a magnetic flux $\Psi_{\rm BH}\sim\pi r_{\rm H}^2B_{\rm H}$, where $J$ is the angular momentum of the BH, $c$ is the light speed, $G$ is the gravitational constant, $B_{\rm H}$ is the strength of the magnetic field at the BH and $r_{\rm H}$ is the radius of the BH horizon.
Then, the total power of the BZ jet is \citep[e.g., ][]{BZ77, TNM11}
\begin{eqnarray}\label{sec2:L_BZ}
L_{\rm BZ}&\sim&\frac{\kappa}{4\pi c}\Omega_{\rm H}^2\Psi_{\rm BH}^2,
\end{eqnarray}
where $\kappa\approx0.05$, the angular frequency of the BH is
\begin{eqnarray}
\Omega_{\rm H}=\frac{a_{\ast}c}{2r_{\rm H}},
\end{eqnarray}
and $a_{\ast}\equiv a/M_{\rm BH}$ is the dimensionless spin parameter.

\underline{Phases I -- III in Figure~\ref{figure:model}:} The rotational energy of the BH formed after the binary NS merger is huge. 
After the inspiral phase of the binary NS merger (phase I), a hypermassive neutron star (HMNS) is formed (phase II), whose gravitational collapse is prevented by differential rotation and thermal pressure \citep[e.g., ][]{BBM13}.
Within the transport timescales of angular momentum and thermal energy ($\lesssim10^{-2}-1$ s), the HMNS eventually collapses to a BH with its surrounding torus (phase III) \citep[e.g., ][]{BBM13, SKKS15}. 
From the numerical simulations, the dimensionless spin parameter of the collapsed BH is $a_{\ast}\sim0.7$ \citep{ST06}. 
Then, the available rotational energy of the BH is 
\begin{eqnarray}
E_{\rm rot}=\left(1-\sqrt{\frac{1+\sqrt{1-a_{\ast}^2}}{2}}\right) M_{\rm BH} c^2 \sim 2\times10^{53}~{\rm erg}, 
\end{eqnarray}
which is enough to explain the total energy of a short GRB. The problem is how to extract the rotational energy up to long timescale, $\sim10^{4}$ s.

A magnetic field is amplified after the merger. 
In pre-merger phase, a non-recycled NS has a dipole magnetic field $B_{\rm NS}\sim10^{12}$ G (phase I). 
Since the radius of NS, $R_{\rm NS}\sim10$ km, is almost equal to that of the BH horizon with the mass $M_{\rm BH}\sim3M_{\odot}$, the strength of the magnetic field at the BH is about $B_{\rm H}\sim B_{\rm NS}$ thanks to the flux conservation.
The corresponding BZ power is too weak to explain the typical observed luminosity of the prompt emission of short GRBs, $\sim10^{49}-10^{51}$ erg s$^{-1}$. 
At least the strength of the magnetic field $B_{\rm H}\sim 10^{14}$ G is required for the BZ process in Equation (\ref{sec2:L_BZ}). 
Therefore, the magnetic field should be amplified by some mechanism such as Kelvin-Helmholtz instability and/or magnetorotational instability \citep[phase II; e.g., ][]{SSKI11, KKSSW14} in matter such as the HMNS and/or the torus.
Recent numerical simulations support such a picture \citep[e.g., ][]{KKSSW14}.

The magnetic field should be expanded to large scale for the BZ process to work.
At the same time, this expansion should be associated with the mass ejection because of the frozen-in condition (phase III). 
The mass ejection is caused by such as winds driven by dynamical interactions \citep[e.g., ][]{Hot+13b, KIS14}, neutrinos \citep[e.g., ][]{Des+09}, magnetic field \citep[e.g., ][]{SSKI11, PRS14} and viscous heating \citep[e.g., ][]{FKMQ14}. 
Although these scenarios do not currently succeed in numerical simulations, the amplification and the expansion of the magnetic field are necessary to explain the observed short GRBs regardless of details.

\underline{Phases IV -- VII in Figure\ref{figure:model}:} 
The central BH accretes mass from the surrounding torus. 
The timescale of the accretion is determined by the viscosity timescale at the inner radius of the torus $t_{\rm vis}\sim0.1$ s \citep[phase IV; e.g., ][]{ZM04}.
It has been widely discussed that the torus accretion could explain the prompt emission of short GRBs. 

In addition, the ejected matter falls back to the BH. 
This is because the ejecta become homologous due to the interaction within ejecta \citep[e.g., ][]{Hot+13b} and the interior part of the ejecta does not exceed the escape velocity (phase V).
The fallback mass is comparable to or larger than that of the escaping ejecta because the inner part of ejecta is more massive \citep{Hot+13b}.
The temporal evolution of the mass accretion rate is described by the following power-law function \citep[e.g., ][]{R07},
\begin{eqnarray}\label{sec2:dotM}
\dot{M}=\frac{2}{3}\frac{M_{\rm f}}{t_{\rm vis}} \left(\frac{t}{t_{\rm vis}}\right)^{-5/3},~~~(t>t_{\rm vis}),
\end{eqnarray}
where $M_{\rm f}\equiv\int_{t_{\rm vis}}^{\infty}\dot{M}dt$ is the total fallback mass.

The fallback enables the long-lasting activities of the BZ jet because the pressure of the fallback matter supports the magnetic flux on the BH, (phases V and VII). As long as the magnetic flux $\Psi_{\rm BH}$ is constant, the BZ power does not directly depend on the mass accretion rate (in Equation~\ref{sec2:L_BZ}) and the light curve becomes plateau-like $L_{\rm BZ}\sim t^0$ \citep{TG15}. 

The important point is that the fallback matter drags the magnetic field line to the BH because of the frozen-in condition and eventually forces the anti-directed magnetic fields to reconnect (phases IV and VI). 
Then, the magnetic flux on the BH $\Psi_{\rm BH}$ as well as the BZ power $L_{\rm BZ}$ decrease. 
In other words, {\it the long-lasting activities require the fallback matter, which itself inevitably leads to the magnetic flux decay. }
This topological argument does not depend on the specific physical processes. 
Different values of the magnetic fields explain the BZ power at different phases, prompt (phase III), extended (phase V) and plateau emission (phase VII) as quantitatively shown in the next section. 
The minimum magnetic flux after the reconnection (phase VII) is determined by the initial flux of a NS before the merger (phase I). 

\underline{Phase VIII in Figure~\ref{figure:model}:} The BH activity ends if the pressure of the fallback matter becomes too small to support the magnetic flux. 
Then, the magnetic field lines escape from the BH (phase VIII). 

\section{LIGHT CURVE}

 \begin{figure}
  \begin{center}
   \includegraphics[width=60mm, angle=270]{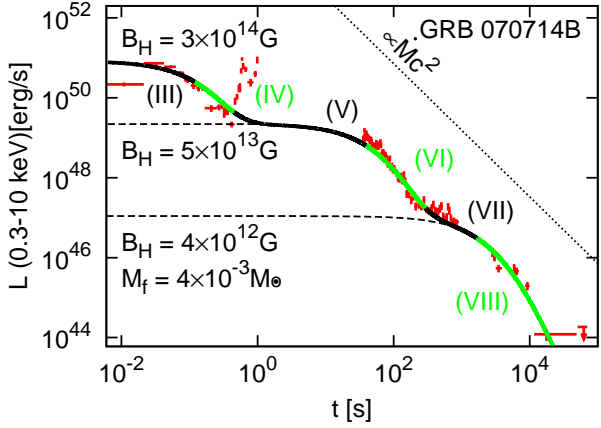}
   \caption{A representative light curve for prompt, extended and plateau emission in our BH model. Observational data of GRB 070714B is obtained from UK {\it Swift} Science Data Centre. Time shown in the horizontal axis denotes the rest-frame time since {\it Swift}/BAT triggers. For the redshift value, we follow \citet{GOWR13}. The number III -- VIII corresponds to the phase in Figure~\ref{figure:model}. For the rebrightening component at $\sim1$ s, we consider flaring activities discussed in Section 4.}
   \label{figure:test}
  \end{center}
 \end{figure}

Figure~\ref{figure:test} shows the theoretical light curve with a thick curve.
Black lines denote the constant luminosity phases (III, V and VII), and light-green lines denote the decay phases (IV, VI and VIII). 
Constant luminosities at three phases (III, V and VII) suggest constant magnetic fields in Equation~(\ref{sec2:L_BZ}); $B_{\rm H}\sim10^{14}$ G at the prompt emission (phase III), $\sim10^{13}$ G at the extended emission (phase V) and $\sim10^{12}$G at the plateau emission (phase VII), respectively.
Here, we assume $a_{\ast}\sim0.7$ \citep{ST06}, so that $r_{\rm H}=(1/2)\left(1+\sqrt{1-a_{\ast}^2}\right)R_{\rm s}\sim0.86R_{\rm s}$, where $R_{\rm s}=2GM_{\rm BH}/c^2$ is the Schwarzschild radius. 
To convert the luminosity from the BZ power $L_{\rm BZ}$ to the observed isotropic luminosity $L$, we take into account the beaming correction \citep[$\theta_{\rm j}^2\sim10^{-3}$;][]{Fong+14} and the radiative efficiency \citep[$\eta\sim0.1$; e.g., ][]{Zha+07}, 
\begin{eqnarray}\label{sec3:L}
L\sim\eta(2/\theta_{\rm j}^2)L_{\rm BZ}\sim10^2L_{\rm BZ}.
\end{eqnarray}
Using Equations (\ref{sec2:L_BZ}) and (\ref{sec3:L}), the strength of the magnetic field is determined by the observed luminosity $L$ as
\begin{eqnarray}\label{sec3:B_H}
B_{\rm H}&\sim&3\times10^{12}~\left(\frac{\eta/\theta_{\rm j}^2}{10^2}\right)^{-1/2}\left(\frac{M_{\rm BH}}{3M_{\odot}}\right)^{-1}\nonumber \\
&&\times\left(\frac{L}{10^{47}{\rm erg~s}^{-1}}\right)^{1/2}~~{\rm G}.
\end{eqnarray}

A characteristic timescale of the BZ jet activity is determined by the pressure balance near the BH.
As the mass accretion rate decreases, the pressure of the fallback matter $p_{\rm f}$ falls short of the magnetic pressure $p_{\rm B}$ \citep[making a magnetically-arrested disk; e.g., ][]{BR76, NIA03}.
Then, the magnetic flux expands the torus and decreases on the BH, leading to the reduction of the BZ power. 
The pressure of the magnetic field is 
\begin{eqnarray}\label{sec3:p_B}
p_{\rm B}=\frac{B_{\rm H}^2}{8\pi}\left(\frac{R}{r_{\rm H}}\right)^{-4}, 
\end{eqnarray}
under the magnetic flux conservation, where $R$ is the radial distance. 
The pressure of the fallback matter is 
\begin{eqnarray}\label{sec3:p_f}
p_{\rm f}=\frac{GM_{\rm BH}\dot{M}}{2\pi R^3v_{\rm R}},
\end{eqnarray}
where $v_{R}$ is the radial velocity. 
For the radial velocity $v_{\rm R}$, we assume $v_{\rm R}\equiv\epsilon v_{\rm ff}$ where $v_{\rm ff}=\sqrt{GM_{\rm BH}/R}$ is the free-fall velocity. 
For the value of $\epsilon$, we adopt $\epsilon\sim10^{-2}$ which is supported by the observations and numerical simulations of the relativistic jets \citep[e.g., ][]{TNM11, ZCST14}. 
Then, the magnetospheric radius $R_{\rm m}$, where the equilibrium point between two pressures $p_{\rm f}$ (Equation \ref{sec3:p_B}) and $p_{\rm B}$ (Equation \ref{sec3:p_f}), is
\begin{eqnarray}\label{sec3:R_m}
\frac{R_{\rm m}}{r_{\rm H}}&\sim&2~\left(\frac{\epsilon}{10^{-2}}\right)^{2/3}\left(\frac{B_{\rm H}}{10^{12}{\rm G}}\right)^{4/3}\nonumber \\
& &\times\left(\frac{M_{\rm BH}}{3M_{\odot}}\right)^{4/3}\left(\frac{\dot{M}}{10^{-11}M_{\odot}~{\rm s}^{-1}}\right)^{-2/3}.
\end{eqnarray}
Using Equation~(\ref{sec3:R_m}) and the temporal evolution of the mass accretion rate (Equation~\ref{sec2:dotM}), we obtain the characteristic timescale of the BZ jet
\begin{eqnarray}\label{sec3:T}
T&\sim&1\times10^4~\left(\frac{\epsilon}{10^{-2}}\right)^{-3/5}\left(\frac{t_{\rm vis}}{0.1{\rm s}}\right)^{2/5}\left(\frac{M_{\rm f}}{10^{-3}M_{\odot}}\right)^{3/5} \nonumber \\
& &\times\left(\frac{M_{\rm BH}}{3M_{\odot}}\right)^{-6/5}\left(\frac{B_{\rm H}}{10^{12}{\rm G}}\right)^{-6/5}~{\rm s}.
\end{eqnarray}
Using Equations~(\ref{sec3:B_H}) and (\ref{sec3:T}), the total fallback mass is determined by the observed luminosity $L$ and duration $T$ as
\begin{eqnarray}\label{sec3:M_f}
M_{\rm f}&\sim&1\times10^{-2}M_{\odot}~\left(\frac{\epsilon}{10^{-2}}\right)\left(\frac{\eta/\theta_{\rm j}^2}{10^2}\right)^{-1}\left(\frac{t_{\rm vis}}{0.1{\rm s}}\right)^{-2/3}\nonumber \\
&&\times\left(\frac{L}{10^{47}{\rm erg~s}^{-1}}\right)\left(\frac{T}{10^4{\rm s}}\right)^{5/3}.
\end{eqnarray}

\begin{figure*}
  \begin{center}
   \includegraphics[width=190mm]{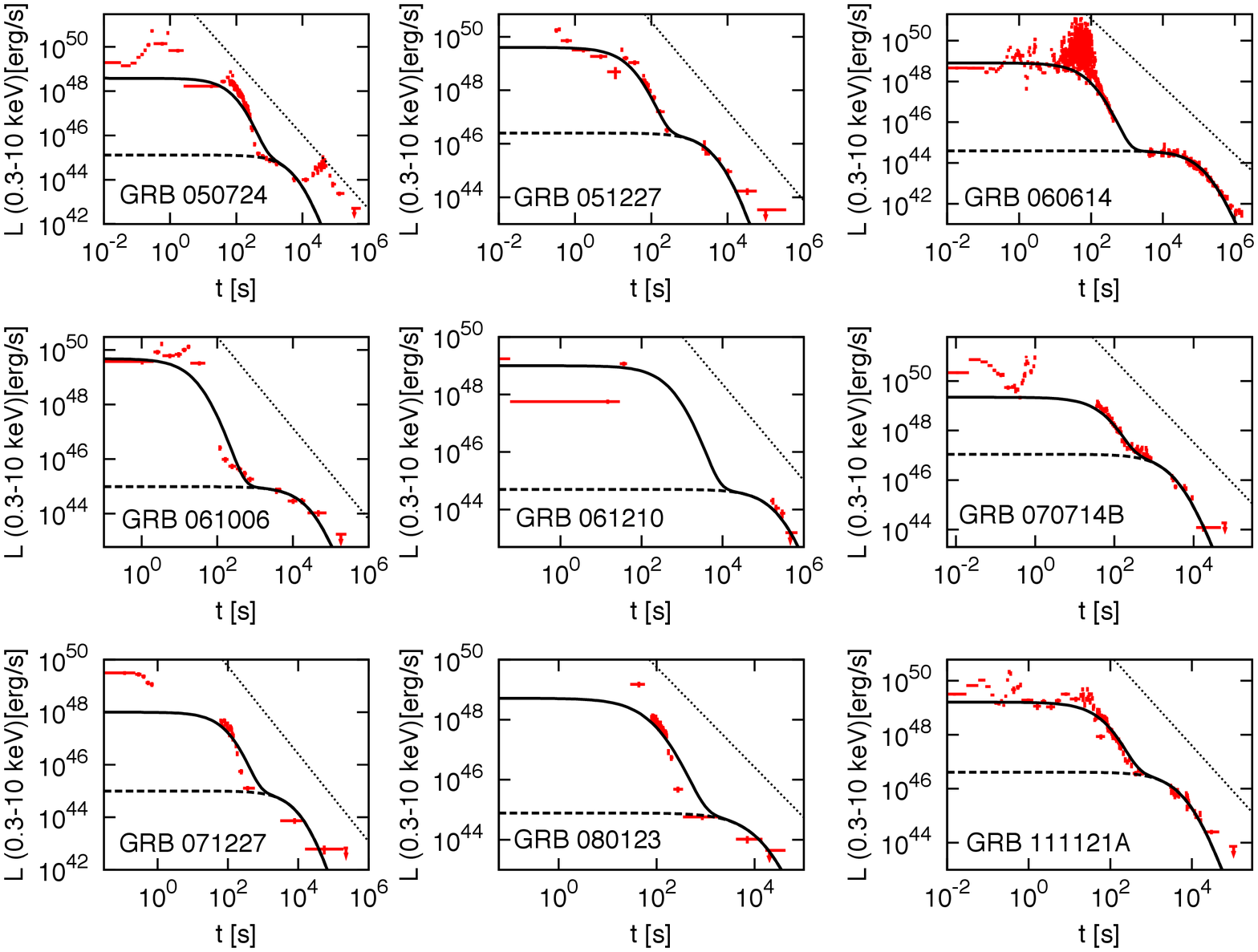}
   \caption{Theoretical light curves for 9 short GRBs used in \citet{GOWR13, GOW14}. 
 For the redshift values, we follow \citet{GOWR13} except for GRB 051227 \citep[$z=0.8$; ][]{DAv+09}.}
   \label{figure:obs}
  \end{center}
 \end{figure*}

After the time $t>T$, the BZ power evolves as $L_{\rm BZ}\propto t^{-40/9}$ derived by the magnetic flux $\Psi_{\rm BH}\propto R_{\rm m}^{-2}$ and the time dependence $R_{\rm m}\propto t^{10/9}$ from Equations~(\ref{sec2:dotM}) and (\ref{sec3:R_m}).
Thus, we model the BZ power as
\begin{eqnarray}\label{sec3:L_BZ}
L_{\rm BZ}\propto\left(1+\frac{t}{T}\right)^{-40/9}.
\end{eqnarray}

The maximum value of the BZ power $L_{\rm BZ}$ is determined by the accretion power $\dot{M}c^2$. 
In fact, using the condition $p_{\rm f}>p_{\rm B}$ and Equations (\ref{sec2:L_BZ}), (\ref{sec3:p_B}) and (\ref{sec3:p_f}), the ratio is $L_{\rm BZ}/(\dot{M}c^2)\lesssim1$. 
For comparison, we plot the mass accretion rate with the beaming correction $(2/\theta_{\rm j}^2)\dot{M}c^2\sim10^3\dot{M}c^2$ in Figures~\ref{figure:test} and \ref{figure:obs} as thin dotted lines. 

Our theoretical light curve is consistent with the observations. 
As an example, we plot the observational data\footnote{http://www.swift.ac.uk/index.php} of short GRB 070714B, which has extended and plateau emission in Figure~\ref{figure:test}. 

In Figure~\ref{figure:obs}, our model is consistent with other 9 short GRBs used in \citet{GOW14}. 
The ratio $L_{\rm BZ}/(\dot{M}c^2)\lesssim1$ is satisfied for all samples. 

\section{DISCUSSION}

\begin{figure}
  \begin{center}
   \includegraphics[width=70mm, angle=270]{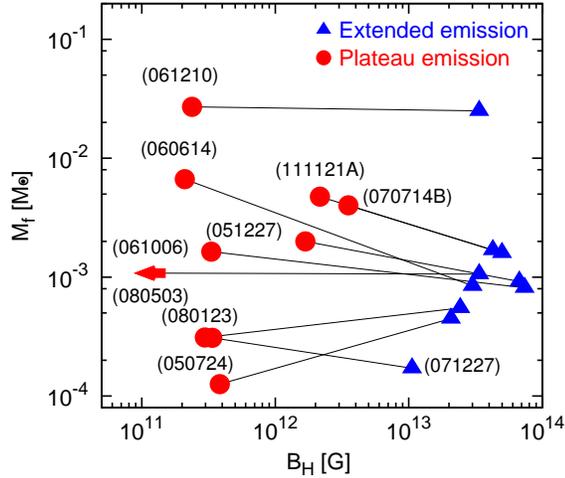}
   \caption{ Estimated total mass of the fallback matter $M_{\rm f}$ and magnetic field $B_{\rm H}$ for extended emission (blue triangles) and plateau emission (red circles). Solid lines connect each event of short GRBs.}
   \label{figure:fallback}
  \end{center}
 \end{figure}

Figure~\ref{figure:fallback} shows the total fallback mass $M_{\rm f}$ and magnetic fields $B_{\rm H}$ at extended (blue triangles) and plateau emission (red circles) evaluated by the observed luminosity $L$ and duration $T$ with Equations (\ref{sec3:B_H}) and (\ref{sec3:M_f}).
The magnetic field at the plateau emission is $B_{\rm H}\sim10^{11}-10^{12}$ G, which is consistent with a typical value of pulsars and the observed value of a non-recycled pulsar, PSR J0737+3039B, in the double pulsar system \citep{Lyne+04}.
For the extended emission, slightly strong magnetic field $B_{\rm H}\sim10^{13}$ G is required to explain the observations.
Different values of the magnetic field between prompt and extended emission may suggest that the magnetic field in the torus formed at the collapse of the HMNS is larger than that in the ejecta. 
On the other hand, the total fallback masses $M_{\rm f}$ for both extended and plateau emission in Figure~\ref{figure:fallback} are consistent with the ejecta mass $M_{\rm f}\sim10^{-4}-10^{-2}M_{\odot}$ obtained from the numerical simulations \citep{Hot+13b} and the observation of a macronova following GRB 130603B \citep[e.g., ][]{Tan+13, BFC13,TNI14,KIT15}.
The fallback masses for the extended and plateau emission are similar in each event, partly supporting our model.

In Figure 4, we also compare our BH model with GRB 080503, which has a bright extended emission and a weak or no plateau emission. The parameters were unobtainable in the magnetar model \citep{GOWR13, GOW14}. We obtain only the upper limit on the magnetic field $B_{\rm H}$ at the plateau phase (red arrow), which are consistent with the other events.

Rapid declines of some light curves may indicate that the magnetic field is decayed by reconnection since Equation (\ref{sec3:L_BZ}) assumes the flux conservation. 
Note that the released energy due to the reconnection is negligible for the extracted energy by the BZ process. 

Some events are accompanied by X-ray flares \citep[e.g., ][]{Mar+11}.
These activities could be explained by the accretion of blob with the same magnetic polarity as $\Psi_{\rm BH}$, the decrease of the jet opening-angle $\theta_{\rm j}$ \citep{MI13} and/or the increase of the radiative efficiency $\eta$. 
After the time $t>T$, the accretion could be episodic, so that the magnetic flux on the BH $\Psi_{\rm BH}$ fluctuates \citep[e.g., ][]{PB03}.
This potentially gives the flaring activities \citep[e.g., ][]{ZP06}.

The extended or plateau emission is only seen in a fraction of the short GRBs. As mentioned in Section 1, the plateau emission could be hidden by the afterglow emission or below the detection limit. Outflows from the accretion disk could also interact with the fallback matter \citep{FKMQ14} and reduce the extended and plateau emissions.

The model in Figure~\ref{figure:model} is applicable to the merger of a NS-BH binary.
The merger produces a BH-torus system with mass ejection \citep[e.g., ][]{KIS13, Fou+15, Kyu+15}.
Difference from the case of the binary NS merger is that a HMNS is not formed in the NS-BH merger case. 
However, the magnetic field amplification in the torus is possible \citep[e.g., ][]{PRS14}. 
Therefore, the phases III -- VIII would be the same for both cases.

The model in Figure~\ref{figure:model} is also applicable to long GRBs since the mass ejection and fallback could also occur with the supernova explosion and/or the central engine activities.
This topic will be discussed in future work.

The long-lasting activities would significantly affect the macronovae \citep{KIT15} since the energy injected to the preceding ejecta suffers from relatively small adiabatic cooling.
This issue will be studied in a separated paper. 

\acknowledgments
We are grateful to the anonymous referee for helpful comments. We would also like to thank K. Asano, K. Kiuchi, K. Kyutoku, T. Nakamura, Y. Sekiguchi, and H. Takami for fruitful discussions. This work is supported by KAKENHI 24103006 (S.K., K.I.), 24000004, 26247042, 26287051 (K.I.).

\end{document}